\documentclass[conference]{IEEEtran}

\usepackage{textcomp}
\usepackage{comment}
\usepackage{ifthen}
\usepackage{subcaption}
\usepackage{booktabs}
\usepackage{footnote}
\usepackage{graphicx}
\usepackage{paralist}
\usepackage{enumitem}
\usepackage{float}
\usepackage{color}
\usepackage{xspace}	
\usepackage{flushend}
\usepackage{balance}
\usepackage{enumitem}
\usepackage{marvosym}
\usepackage{tikz}

\def\BibTeX{{\rm B\kern-.05em{\sc i\kern-.025em b}\kern-.08em
    T\kern-.1667em\lower.7ex\hbox{E}\kern-.125emX}}

\newboolean{authnotes}

\ifthenelse{\boolean{authnotes}}
{
	\newcommand{\amy}[1]{\footnote{{\bf Amy: #1}}}
	\newcommand{\rajeev}[1]{\footnote{{\bf Rajeev: #1}}}
	\newcommand{\michele}[1]{\footnote{{\bf Michele: #1}}}
}
{
	\newcommand{\amy}[1]{}
	\newcommand{\rajeev}[1]{}	
	\newcommand{\michele}[1]{}
}

\newcommand{\fakeparagraphnodot}[1]{\vspace{0mm}\noindent\textbf{#1}}
\newcommand{\fakeparagraph}[1]{\fakeparagraphnodot{#1}}

\newcommand{\kos}{KRATOS\xspace}

\begin{document}

\title{KRATOS: An Open Source Hardware-Software Platform for Rapid Research in LPWANs}

\author{\IEEEauthorblockN{
		\IEEEauthorrefmark{1}Rajeev~Piyare, 
		\IEEEauthorrefmark{1}Amy~L.~Murphy, 
		\IEEEauthorrefmark{2}Michele~Magno, and  
		\IEEEauthorrefmark{2}Luca~Benini}
	
	\IEEEauthorblockA{
		\IEEEauthorrefmark{1}Fondazione Bruno Kessler, Trento, Italy \{piyare, murphy\}@fbk.eu} 
	\IEEEauthorrefmark{2}Integrated Systems Laboratory, ETH Z\"{u}rich \{michele.magno, lbenini\}@iis.ee.ethz.ch
}

\maketitle

\begin{abstract}
Long-range (LoRa) radio technologies have recently gained momentum in the IoT landscape, allowing low-power communications over distances up to several kilometers. As a result, more and more LoRa networks are being deployed.  However, commercially available LoRa devices are expensive and propriety, creating  a barrier to entry and possibly slowing down developments and deployments of novel applications.
Using open-source hardware and software platforms  would allow more developers to test and build intelligent devices resulting in a better overall development ecosystem, lower barriers to entry, and rapid growth in the number of IoT applications. Toward this goal, this paper presents the design, implementation, and evaluation of \emph{\kos}, a low-cost LoRa platform running ContikiOS. Both, our hardware and software designs are released as an open-source to the research community.
 
\end{abstract}



\section{Introduction}
\label{sec:introduction}

The Internet of Things is the new Internet frontier providing networks between smart physical objects or ``Things" embedded with various sensors. IoT offers connectivity and services enabling applications spanning from industrial to consumer sectors where sensors continuously monitor and detect physical parameters like energy usage, temperature, indoor and outdoor air quality, leakage, and intrusion detection. 

A founding pillar of the IoT concept is the availability of low-cost low-power devices with wireless technologies providing both sensing and actuation capabilities.
Long-range (LoRa) radio technologies have recently gained momentum in the IoT landscape. These technologies operate in the Sub-GHz bands, allowing low-power communications over distances up to several kilometers. As a result, LoRa is now been considered a candidate radio technology for many low-power wide area network (LPWAN) applications, especially those that require extended coverage such as citywide sensing, wildlife, or remote infrastructure monitoring~\cite{raza_ieee}. 

Nevertheless, commercially available LoRa devices such as sensor nodes and gateways are expensive and propriety, creating  a barrier to entry and possibly slowing down developments and deployments of novel applications. This trend is, however, changing with the availability of open-embedded platforms and a wide selection of the commercial off-the-shelf (COTS) components that can be used to realize prototypes and solutions quickly~\cite{openHW}. As our first contribution, we provide an open source hardware platform, \kos  for rapid prototyping and testing of LoRa networks using COTS components. The developed platform compares favorably w.r.t other commercially available  devices in terms of performance, power, and cost as shown in Table~\ref{tab:1}.

One important aspect of battery operated IoT devices is low-power consumption and extended battery life up to several years. Design of low-power hardware alone is not sufficient to fully realize energy-efficient systems. A hardware-software co-design is requisite that ensures that the complete system is compatible with the standards providing reliable performance while keeping power consumption at check. The availability of novel energy-efficient networking protocols for LoRa networks is lacking as well, compared to its IEEE 802.15.4 wireless counterparts for which a vast majority of platforms and software networking stacks exist. 

\begin{table*}[!t]
	\centering
	\caption{Qualitative comparison of different LoRa based sensor motes.}
	\label{tab:1}
	\begin{tabular}{@{}lcccccc@{}}
		\toprule
		\textbf{Features} & LoRa Fabian & SODAQ ONE & NetBlocks & WeMos-Lora & WizziKit & \textbf{\kos} \\ \midrule
		MCU & ATmega328 & ATSAMD21G18 & STM32L151 & ESP8266 & STM32L0 & MSP430FR5969 \\ \midrule
		Flash size (KB) & 32 & 256 & 256 & x & 192 & 64 \\ \midrule
		On-board sensors & x & 3 & x & x & 7 & 8 \\ \midrule
		Operating System  & x & x & x & x & x & ContikiOS \\ \midrule
		Networking stack & partially & x & x & x & partially & full \\ \midrule
		Open hardware/software & no/yes & no/yes & no/yes & yes & no/yes & yes \\ \midrule
		Energy harvesting & no & no & no & no & no & yes \\ \midrule
		On-board WuRX & no & no & no & no & no & yes \\ \midrule
		Cost (\$) & 115 & 109 & 72 & x & 102 & $<$100 \\ \bottomrule
	\end{tabular}
\end{table*}

Over the past decade, the WSN community has produced a vast number of networking protocols, raising the question of how much of this research outcome can be re-used for LoRa based networks  without reinventing the wheel. To address this question, the Contiki operating system~\cite{contiki} can be regarded as a key enabler, as it has received wide adoption both from academia and industry, making the OS increasingly interesting for IoT applications. ContikiOS supports various standardized networking protocols such as Collection Tree Protocol, IPv6 over Low-Power Wireless Personal Area Networks (6LoWPAN), and IPv6 Routing Protocol for Low-Power and Lossy Networks (RPL). The barrier, however, consists in the inability to directly run ContikiOS atop LoRa platforms. Although, there have been some efforts made to port Contiki for LoRa radios~\cite{IPv6lora, 8115756}, however, none of these implementations are publicly available. In this paper, we present an open-source ContikiOS stack for facilitating development and testing of novel LoRa based protocols using \kos platform.  This will also open up possibilities to experiment asynchronous and synchronous protocols provided by Contiki for star and multi-hop LoRa networks.

In summary, the contributions of this work are as follows:
\begin{enumerate}[label=(\roman*)]
	\item design of a multi-sensor long-range wireless platform for the IoT. To operate autonomously and  in a very low-power state, the sensor mote has energy harvesting capabilities together with an on-board ultra-low power wake-up receiver for on-demand communication. To the authors best knowledge, this work provides the first open-source hardware design of the wake-up receiver to the research community (\S\ref{sec:platform}).
	\item ContikiOS port for the developed mote that includes:
		\begin{inparaenum}[\em i)]
			\item MSP430FR5969 microcontroller from TI, extending the Contiki hardware support list. This will enable to easily interface other 802.15.4 radios that are already supported by Contiki such as CC1200, CC2420, CC2520.
			
			\item SX1276 Semtech LoRa radio transceiver to enable long-range application designs (\S\ref{sec:contiki}).
		\end{inparaenum}
	\item \emph{open-source }: both the hardware and the software design of the proposed platform and its tool-chain bundled as \kos\footnote{The open-source hardware and software release of the \kos framework can be found at: \\
		 https://contikios4lora.github.io/contikios-lora/} are released to the research community in contrast to the other commercial platforms.
\end{enumerate}

We argue that our open source \kos framework can become an enabler for research revolving around long-range radios as well as help accelerate the development and testing of new LoRa based prototypes and products without building from scratch.


\section{KRATOS Design }

There has been much work on the general idea of open source wireless sensor motes. While there is a similarity between previous efforts and ours at a high-level, the crucial difference is that \kos is  a multi-radio platform that operates at ultra low-power regimes between 1.83$\mu W$ in sleep state and 240$mW$ when transmitting at +14dBm,  and can be switched almost instantly using the wake-up receiver for on-demand communication. The prototype of the wireless multi-sensor dual-radio platform is shown in Fig.~\ref{fig:prototype}. 
The overall cost of our platform is estimated to be less than \$100 for a quantity of 100 nodes and less than \$50 for 10000 nodes.

\subsection{Hardware}
\label{sec:platform}
Next, we present the hardware features of the new mote followed by the operating system that runs on it. The LoRa mote includes the following main components:

\begin{figure}[!b]
	\centering
	\includegraphics[width=0.9\linewidth]{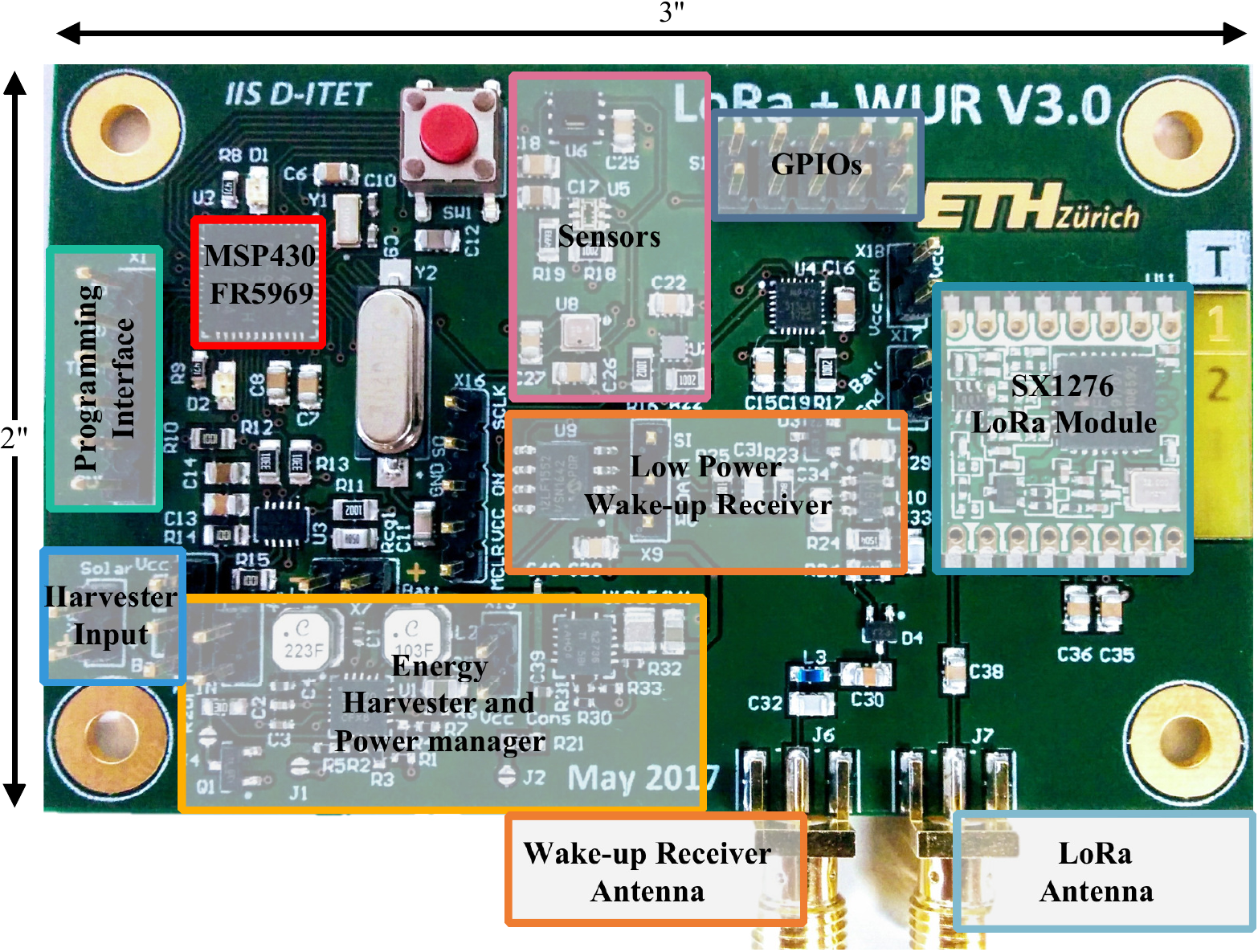}
	\caption{The hardware architecture of our LoRa sensor mote: the MSP430FR5969 runs applications and the SX1276 manages long-range communication.}
	\label{fig:prototype}
	\vspace*{-3.0ex}
\end{figure}

\fakeparagraph{Processor.}
The platform is built around a 16-bit MSP430FR5969 microcontroller from Texas Instruments. This particular microcontroller was chosen due to its ultra-low power consumption in sleep mode (0.3$\mu A$ in LPM4), very fast wake-up time (7$\mu s$ from LPM4 to active mode) and presence of an on-board FRAM. This non-volatile memory allows savings of all register values without loss of states in the events of power disruptions or energy depletion. The MSP430FR5969 is the core of our system as it runs the applications to control the main transceiver, sensors, and the power management policies.

\fakeparagraph{RF front-end.}
To have a flexible wireless sensor platform for future research opportunities, we integrated a multi-purpose SX1276 wireless transceiver from Semtech Corporation. The MSP430 communicates with the radio chip over the SPI bus. The transceiver allows short and long-range communication by switching between different modulations techniques such as (G)FSK and OOK, as well as the LoRa physical layer using a single radio module. Sensor mote also hosts a wake-up receiver, therefore, to use SX1276 as a wake-up transmitter, it is configured for transmission using OOK modulation where the information is sent using `1's and `0's. An OOK 1 sub-bit is produced by transmitting a large amplitude carrier while an OOK 0 sub-bit is produced by sending nothing i.e., the transmitter is turned off. Thus, allowing the system to save on transmit power when (not) sending ‘0’s. 

\fakeparagraph{Wake-up receiver (WuRX).}
Together with the multi-purpose SX1276 wireless transceiver, each sensor mote is also equipped with an ultra-low power WuRX. The WuRX is designed such that its power consumption is in order of micro-watts, i.e., 1.8$\mu W$ in standby mode when listening for the signal. The impedance matching circuitry has been tuned for the 868 MHz band with receiver sensitivity of -50 dBm. WuRX also incorporates a passive OOK demodulator coupled to an 8-bit ultra-low power PIC12LF1552 MCU from Microchip for decoding an address embedded in the RF carrier for selective triggering and for preventing false wake-ups. We have leveraged the single digital output pin (GPIO) from PIC MCU as a DATA line that is interfaced to the main processor  for triggering the mote. The WuRX supports a maximum bit rate of 1~kbps. 

\fakeparagraph{On-board sensors.}
To be suitable and flexible for sensing applications, the platform has been embedded with various analog and digital sensors: (i) light, (ii) humidity, (iii) temperature, (iv) barometer, (v) infrared sensor, and (vi) an inertial measurement unit (IMU). Most of these sensors are interfaced to the main microcontroller through an $I^{2}C$ bus for data acquisition.

\fakeparagraph{Power management unit (PMU).}
The power for the developed platform can be either supplied with Lithium-ion batteries or energy harvesting sources. The PMU is designed using the BQ25570 low-power harvester IC that integrates maximum power point tracking with the energy efficiency of $\approx$90\%. This IC has been chosen due to its very low start-up voltage (≥100 mV) making it suitable for indoor energy harvesting from solar panels or thermometric transducers. Finally, the IC can recharge two on-board super-capacitors and Li-Ion battery with the same efficiency.


\subsection{Operating System for the IoT}
\label{sec:contiki}
The first step toward enabling the full networking stack for LoRa is to get the ContikiOS  and firmware tool-chains  ported. In this section, we describe the details of our Contiki port for the developed sensor mote.

\begin{figure}[t!]
	\centering
	\includegraphics[scale=0.5]{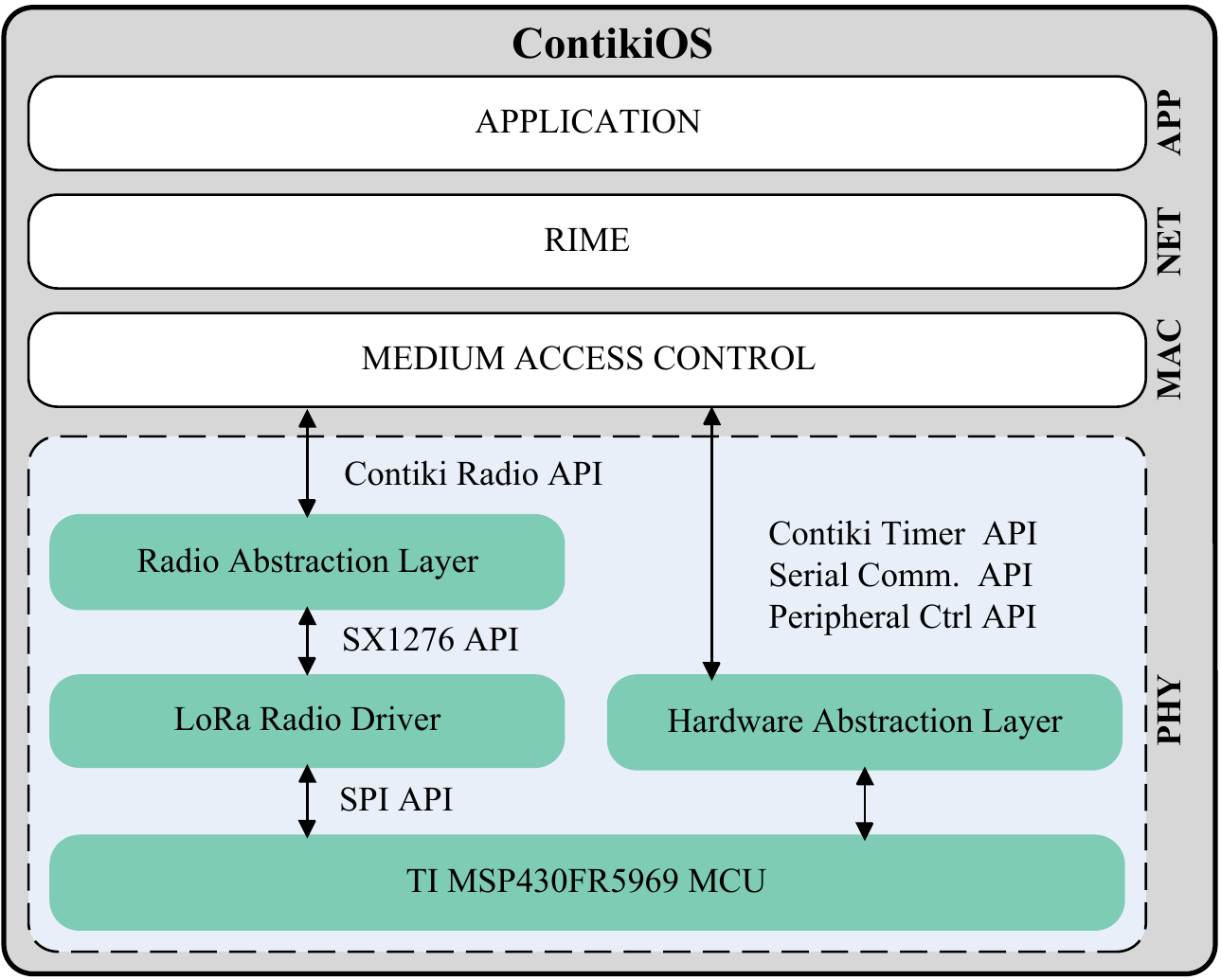}
	\caption{ContikiOS LoRa  software stack}
	\label{fig:contiki-port}
	\vspace*{-3.0ex}
\end{figure}

\fakeparagraph{Porting ContikiOS.}
In Contiki, providing  a support for a new hardware architecture means implementing hardware abstraction layer for the processor and its various peripherals. Our port introduces such modules for the essential hardware of the LoRa platform, including startup code for the MSP MCU, timers, the SPI bus, serial line input and output (UART), and the radio. Fig.~\ref{fig:contiki-port} depicts the high-level view of the software architecture with the green blocks representing the main components that we implemented.

\fakeparagraph{Hardware Abstraction Layer (HAL).}
As an essential part of this work, we ported the ContikiOS to the TI MSP430FR5xx MCU series, which was unavailable
at the time of writing this paper.  Before the LoRa radio driver could be ported, we had to provide the port for the microcontroller unit.  This port relies on low-level libraries provided by the manufacturer of the MCU. The HAL is a thin layer
that manages the low-level Contiki API calls to the MCU such as interrupt handling, power management, and watchdog timers. The HAL also allows configuring the MCU and its external peripherals to be used by Contiki such as buttons, LEDs, sensors etc.

\fakeparagraph{Radio Abstraction Layer (RAL).} 
The LoRa radio driver is based on the reference library provided by Semtech for Cortex-M3 processors~\cite{semtech}. 
The Semtech low-level radio driver provides all necessary functions to operate the SX1276 radio chip. However, ContikiOS relies on a RAL with its custom APIs and a set of functional requirements that all radio drivers should adhere to be compliant with the upper stacks of the OS. Therefore, we created an adaptation layer translating the Contiki API calls into the SX1276 API calls. We customize the original driver from Semtech to make it compatible for the MSP430 family. RAL is a platform-specific radio interface that (i) initializes the LoRa transceiver, (ii) configures the radio parameters (bandwidth, spreading factor, coding rate, power, frequency), (iii) manages the transmission and reception of data packets, (iv) performs channel checks, and (v) controls  the radio chip. 

\fakeparagraph{Support for other LoRa chipsets and MCUs.} 
Our current implementation for the SX1276 radio chip can be easily extended to other LoRa chip-sets such as SX1272/77/78/79 as the radio physical layer is essentially cross-platform. In addition, other MCU families can easily access the radio as the port relies on a standard SPI interface to communicate with the chipset and utilizes  GPIOs to enable/disable radio interrupts.

\section{Evaluation}
\label{sec:eval}
We ran two different applications (i) communication range using Rime networking stack (ii) and hardware power profiling for validating our \kos framework that includes the sensor mote and the ContikiOS.

\begin{figure*}[!t]
	\begin{minipage}[b]{0.3\textwidth}
		\centering
		\includegraphics[width=0.9\textwidth, height=0.15\textheight]{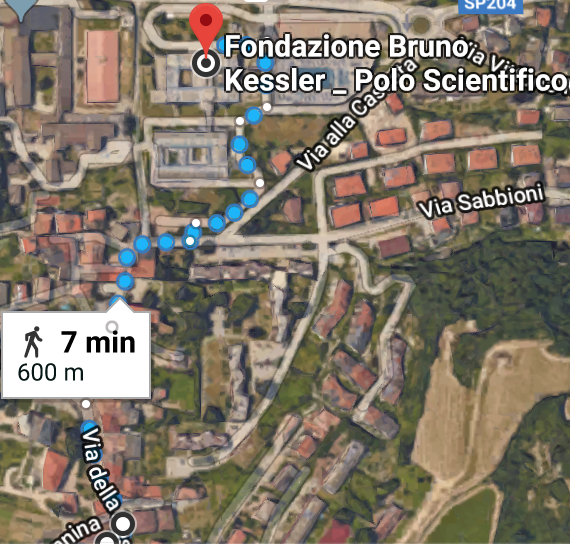}
		\caption{Route taken to evaluate the mote communication range using Rime unicast application.}
	\end{minipage}
	\hfill
	\begin{minipage}[b]{0.34\textwidth}
		\centering
		\includegraphics[width=1.0\textwidth]{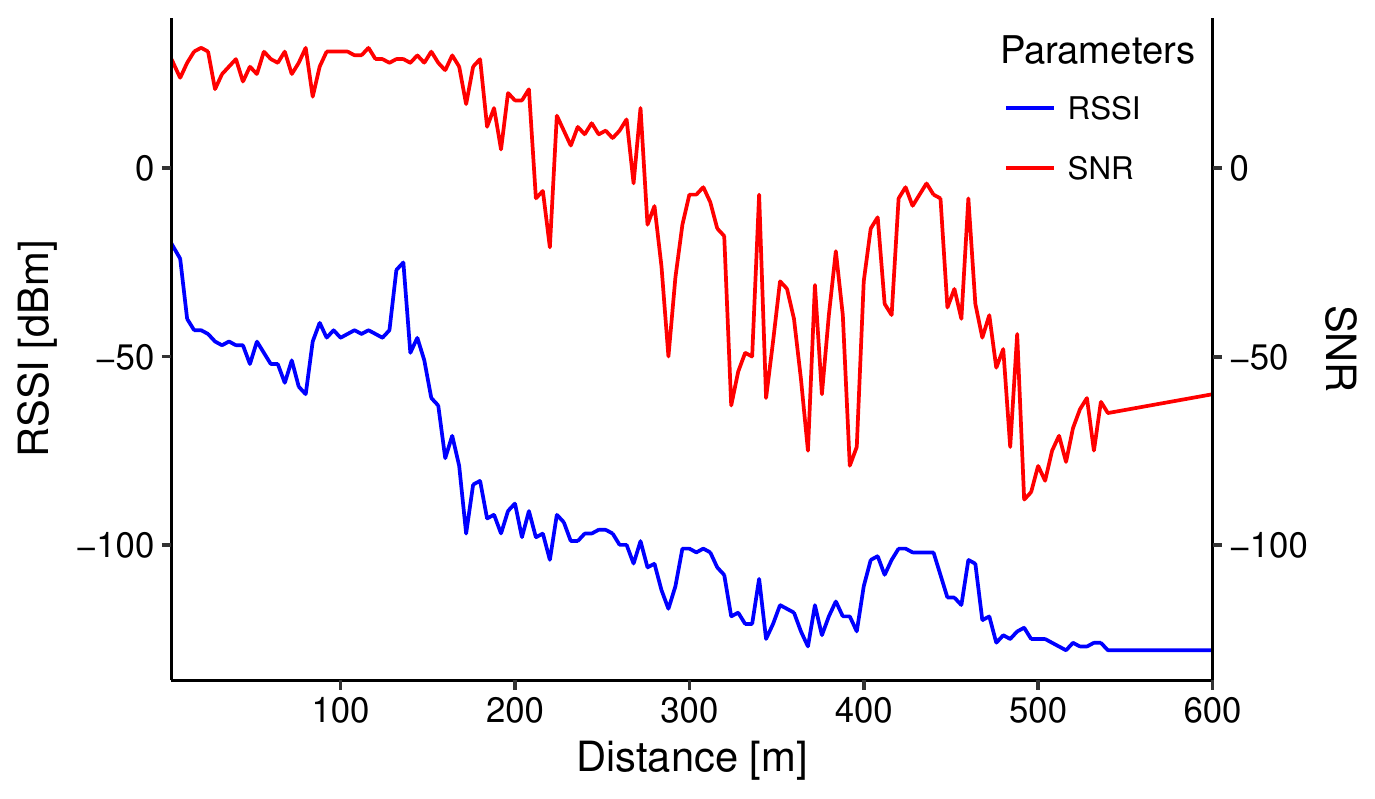}
		\caption{Received signal strength  and signal to noise ratio measurement of the LoRa packet along the route shown in Fig.~3.}
	\end{minipage}
	\hfill
	\begin{minipage}[b]{0.34\textwidth}
		\centering
		\includegraphics[width=1.0\textwidth]{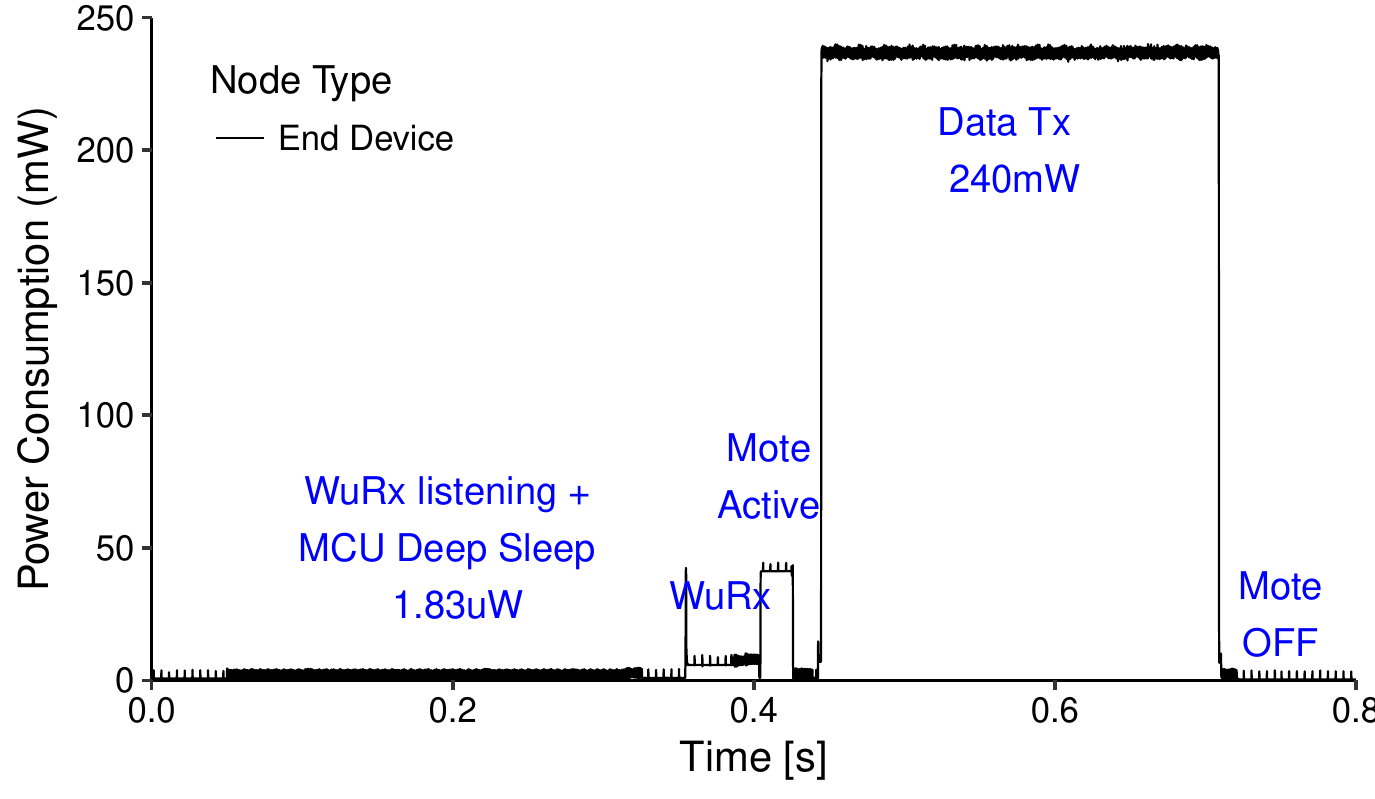}
		\caption{In-lab measurement of the power consumption of the mote under different operation modes.}
	\end{minipage}
\vspace*{-3.0ex}
\end{figure*}

\subsection{Network coverage}

We conduct an experiment to evaluate the communication range of our designed LoRa prototype. Since our platform port is compliant with the upper stack of the ContikiOS, this allows us to run unmodified networking protocol,  \emph{Rime} atop LoRa physical layer as illustrated in Fig.~\ref{fig:contiki-port}. Rime is a light-weight highly-modular stack in ContikiOS providing services such as dissemination, point-to-point routing, multipoint-to-point data collection, and multi-hop communication~\cite{Dunkels:2007}. It also supports uIPv6 stack comprising 6LoWPAN, RPL IPV6 routing, TCP, UDP, and CoAP. 

We ran a simple Rime unicast application where the sensor mote generates and sends a 16B data packet at every 10s to the base station (BS). We place the BS connected to a laptop near the  office window in the FBK building. The sensor mote was carried along the route shown in Fig.~3 to estimate the coverage range of the mote.  Along the data packet, we also collected the received signal strength (RSSI) and the signal to noise ratio (SNR) obtained from the radio abstraction layer in Contiki and is plotted in Fig.~4. The LoRa radio was configured with the transmit power of +14~dBm, bandwidth of 500~kHz, spreading factor of 12, and a coding rate of $4/6$. 

It is worth mentioning that the evaluated route is slightly hilly, with an 8-10m height difference from the location of the BS to the end of the route. Along the route, there are a few vegetation and buildings preventing LOS communication. Nevertheless, at the distance of 600~m, the BS was still able to receive the data packets over 95\% reliability with RSSI $>$-120 dBm. The minimal RSSI required to successfully communicate with the BS is -140~dBm~\cite{sx1276}, indicating that in our measurement, the maximal communication range has not been reached and we aim to explore this in the future. The SNR values in the evaluation range from 32 to -70. The negative SNR value is due to the non-LOS  communication. 

\subsection{Hardware micro-benchmarking}
Next, we conduct the power profiling of the whole prototype in different modes. In WSN, various power management techniques can be used to achieve low-power operation. This involves putting components like MCU and radio transceiver in an ultra low-power state when not used by the applications to reduce power consumption. Wake-up receivers are a novel hardware solution to achieve low-power, asynchronous communication among nodes~\cite{rajeevlcn}. For benchmarking the prototype, we exploit the wake-up receiver technique. In this evaluation, the MSP430 and the LoRa transceiver spends most of its duty cycle in a low-power deep sleep. The external interrupt line is enabled that triggers the MSP430 out of sleep mode when a valid wake-up signal is received over the wake-up receiver interface. Once the MCU is awake, it turns on the main transceiver for exchanging data. For data collection, we utilize the single-hop unicast primitive provided by the Rime stack.
The power consumed by the mote in different modes is illustrated in Fig.~5. When the MCU is in a sleep state with the wake-up receiver actively listening, the mote  draws 1.83~$\mu W$ of power. This consumption rises to  284~$\mu W$ when the wake-up receiver is decoding the node address. The LoRa transmission at +14~dBm consumes 240~$mW$ and 50~$mW$ in reception, respectively.



\section{Conclusions}
\label{sec:conclusions}
To narrow the barrier between testing and deployment of LoRa networks, this paper presents the design, implementation, and evaluation of a low-cost LoRa platform. To use our platform as a starter kit for research and teaching LoRa technology, we also ported the ContikiOS. This will allow researchers to design and explore energy efficient protocols for LoRa networks. Both, our hardware and software designs are released as an open-source framework, \kos.  We encourage the community to download our code, fabricate LoRa mote, and write applications. We envision that \kos can be used to accelerate research on LoRa networks for large-scale deployments without building prototypes and products from scratch.


\bibliographystyle{IEEEtran}
\bibliography{bib}

\begin{thebibliography}{1}
\providecommand{\url}[1]{#1}
\csname url@samestyle\endcsname
\providecommand{\newblock}{\relax}
\providecommand{\bibinfo}[2]{#2}
\providecommand{\BIBentrySTDinterwordspacing}{\spaceskip=0pt\relax}
\providecommand{\BIBentryALTinterwordstretchfactor}{4}
\providecommand{\BIBentryALTinterwordspacing}{\spaceskip=\fontdimen2\font plus
\BIBentryALTinterwordstretchfactor\fontdimen3\font minus
  \fontdimen4\font\relax}
\providecommand{\BIBforeignlanguage}[2]{{%
\expandafter\ifx\csname l@#1\endcsname\relax
\typeout{** WARNING: IEEEtran.bst: No hyphenation pattern has been}%
\typeout{** loaded for the language `#1'. Using the pattern for}%
\typeout{** the default language instead.}%
\else
\language=\csname l@#1\endcsname
\fi
#2}}
\providecommand{\BIBdecl}{\relax}
\BIBdecl

\bibitem{raza_ieee}
U.~Raza, P.~Kulkarni, and M.~Sooriyabandara, ``{Low Power Wide Area Networks:
  An Overview},'' \emph{IEEE Communications Surveys \& Tutorials}, vol.~19,
  no.~2, pp. 855--873, Secondquarter 2017.

\bibitem{openHW}
R.~Fisher, L.~Ledwaba, G.~Hancke, and C.~Kruger, ``{Open Hardware: A Role to
  Play in Wireless Sensor Networks?}'' \emph{Sensors}, vol.~15, no.~3, pp.
  6818--6844, 2015.

\bibitem{contiki}
A.~Dunkels, B.~Gronvall, and T.~Voigt, ``{Contiki-a Lightweight and Flexible
  Operating System for Tiny Networked Sensors},'' in \emph{Proceedings of the
  29th Annual IEEE LCN}, 2004, pp. 455--462.

\bibitem{IPv6lora}
S.~Thielemans, M.~Bezunartea, and K.~Steenhaut, ``{Establishing transparent
  IPv6 communication on LoRa based low power wide area networks},'' in
  \emph{Wireless Telecommunications Symposium}, April 2017, pp. 1--6.

\bibitem{8115756}
B.~Sartori, S.~Thielemans, M.~Bezunartea, A.~Braeken, and K.~Steenhaut,
  ``{Enabling RPL multihop communications based on LoRa},'' in \emph{IEEE
  WiMob}, Oct 2017, pp. 1--8.

\bibitem{semtech}
\BIBentryALTinterwordspacing
Semtech, \emph{{LoRaWAN Endpoint Stack Implementation}}, 2017. [Online].
  Available: \url{https://github.com/Lora-net/LoRaMac-node}
\BIBentrySTDinterwordspacing

\bibitem{Dunkels:2007}
A.~Dunkels, F.~\"{O}sterlind, and Z.~He, ``{An Adaptive Communication
  Architecture for Wireless Sensor Networks},'' in \emph{Proceedings of the ACM
  SenSys}, NY, USA, 2007, pp. 335--349.

\bibitem{sx1276}
\emph{{SX1276/77/78 LoRa Modem}}, Semtech Corporation, July 2017, rev.0.0.

\bibitem{rajeevlcn}
R.~Piyare, A.~L. Murphy, P.~Tosato, and D.~Brunelli, ``{Plug into a Plant:
  Using a Plant Microbial Fuel Cell and a Wake-Up Radio for an Energy Neutral
  Sensing System},'' in \emph{IEEE 42nd Conference on Local Computer Networks
  Workshops}, Oct 2017, pp. 18--25.

\end{thebibliography}

\end{document}